\documentclass[a4paper,11pt]{article}
\usepackage{posPre}

\title{{\footnotesize 
DESY 22--117, DO-TH 22/18, RISC Report Series 22-08, SAGEX-22-28-E}
\\[0.4cm]
Computer Algebra and Hypergeometric Structures for Feynman Integrals}

\ShortTitle{Computer Algebra and Hypergeometric Structures for Feynman Integrals}

\author[a]{Johannes Bl\"umlein}
\author[a]{Marco Saragnese}
\author*[b]{Carsten Schneider}

\affiliation[a]{Deutsches Elektronen--Synchrotron, DESY, \\
	Platanenallee 6, D--15738 Zeuthen, Germany.}

\affiliation[b]{Johannes Kepler University Linz, Research Institute for Symbolic Computation (RISC),\\ 
	Altenberger Stra\ss{}e 69, 4040 Linz, Austria}

\emailAdd{Johannes.Bluemlein@desy.de}
\emailAdd{Marco.Saragnese@desy.de}
\emailAdd{Carsten.Schneider@risc.jku.at}

\abstract{We present recent computer algebra methods that support the calculations of (multivariate) series solutions for (certain coupled systems of partial) linear differential equations. The summand of the series solutions may be built by hypergeometric products and more generally by indefinite nested sums defined over such products.  Special cases are hypergeometric structures such as Appell-functions or generalizations of them that arise frequently when dealing with parameter Feynman integrals.}

\FullConference{
  Loops and Legs in Quantum Field Theory - LL2022,\\
  25-30 April, 2022\\
 Ettal, Germany
}

\newcommand{\NN}{\mathbb{N}}
\newcommand{\ZZ}{\mathbb{Z}}
\newcommand{\QQ}{\mathbb{Q}}
\newcommand{\KK}{\mathbb{K}}

\newcommand{\ep}{\varepsilon}

\newcommand*\pFqskip{8mu}
\catcode`,\active
\newcommand*\pFq{\begingroup
	\catcode`\,\active
	\def ,{\mskip\pFqskip\relax}%
	\dopFq
}
\catcode`\,12
\def\dopFq#1#2#3#4#5{%
	{}_{#1}F_{#2}\biggl[\genfrac..{0pt}{}{#3}{#4};#5\biggr]%
	\endgroup
}

\begin{document}
\maketitle

\section{Introduction}

For certain classes of topologies Feynman parameter integrals in terms
of the dimensional parameter $\ep = D - 4$, with $D$ the dimension of 
space--time can be represented
in terms of higher transcendental functions of the hypergeometric type, cf.~e.g.~\cite{HAMBERG,Davydychev:2003mv,Bierenbaum:2007qe,Kalmykov:2020cqz}. 
This includes, e.g., general
off--shell representations at the one--loop level for multi--leg diagrams, 
cf.~e.g.~\cite{Boos:1990rg,Fleischer:2003rm,Watanabe:2013ova,Bluemlein:2017rbi,Phan:2018cnz} and at the two- and 
three--loop level for various scattering processes, 
cf.~e.g.~\cite{Anastasiou:1999ui,Anastasiou:1999cx,Bauberger:1994nk,Ablinger:2012qm}.
Such representations of higher transcendental function representations, covering as special cases hypergeometric functions or Appell functions
\cite{APPELL,Kalmykov:2020cqz,HypStructure}
and their variations, are often preferential: they provide further inside
in the classification of the arising functions. In particular, one can expand the summands in $\ep$ and can apply symbolic methods~\cite{EpExpansionMethods} and in particular symbolic summation tools, see~\cite{SymbolicSum,HypStructure}, to obtain the coefficients of special functions in terms of iterative sums~\cite{Vermaseren:1998uu,Blumlein:1998if,Moch:02,Ablinger:2013cf,Ablinger:2011te,Ablinger:2014bra,EpExpansionMethods,
Ablinger:2018brx,Ablinger:2020snj} or integrals~\cite{Remiddi:1999ew,Moch:02,Ablinger:2013cf,Ablinger:2011te,Ablinger:2014bra,Ablinger:2018brx,Ablinger:2020snj}

In most cases such representations cannot be extracted directly from the given Feynman parameter integrals due to their complicated mathematical structure. In such instances one often succeeds in computing linear difference equations in a free discrete parameter (often the Mellin variable $n$) or linear differential equation in a continuous variable $z$ (often in its $z$-space representation).
One successful tool is the Almkvist-Zeilberger algorithm~\cite{AZ:06,Ablinger:21} to compute from a given integral (with an hyperexponential integrand) a linear difference or differential equation. Another tactic is to apply successively Newton's binomial theorem and Mellin--Barnes decompositions on 
the integrand, implemented in different packages 
\cite{MBTechniques}.  Applying afterwards the residue theorem produces definite multiple sums and one can utilize symbolic summation tools~\cite{WZ,SymbolicSum} in order to compute linear differential or difference equations in the free parameters. Finally, a highly successful approach are IBP methods~\cite{IBP} that enable one to represent such integrals in terms of master integrals which themselves can be represented as solutions of coupled systems of linear differential equations. Utilizing then uncoupling methods~\cite{UncouplingMethods} lead again to linear differential equations that contain the desired parameter Feynman integrals as solutions. For further details on these and related technologies we refer, e.g., to~\cite{SAGEX} and references therein.

In Section~\ref{Sec:REDE}  we summarize the available tools to find solutions in terms of iterative sums or integrals for a given linear difference or differential equation. Utilizing these tools we will illustrate useful methods to find hypergeometric series solutions and generalizations of them for ordinary linear differential equations (Section~\ref{Sec:OrdDE}), partial linear differential equations (Section~\ref{Sec:PartialDE}) and special coupled systems of partial linear differential equations (Section~\ref{Sec:SystemPartialDE}). We conclude the article in Section~\ref{Sec:Conclusion}.

\section{Solving difference or differential equations in terms of iterative sums or integrals}\label{Sec:REDE}

As described in the introduction one can use, e.g., symbolic integration and summation tools or IBP methods to compute linear difference or differential equations that contain the parameter Feynman integrals (or parts of them) as solutions. 

Given such a linear difference equation, one may use the summation package~\texttt{Sigma}~\cite{dAlembert,SymbolicSum,SigmaSolver} that enables one to solve the following problem.

\medskip

\noindent\textbf{GIVEN} a recurrence
		$$a_0(n)F(n)+\dots+a_{\delta}(n)F(n+\delta)=h(n)$$
where the coefficients $a_0(n),\dots,a_{\delta}(n)$ and the inhomogeneous part $h(n)$ are expressions in terms of indefinite nested sums defined over hypergeometric products, together with initial values, say\footnote{In the following we assume that $\KK$ is an appropriate computable field of characteristic $0$ that contains all the relevant constants for our concrete calculations.} $F(0),\dots,F(\delta-1)\in\KK$.\\
\noindent\textbf{DECIDE} constructively if $F(n)$ can be expressed in terms of indefinite nested sums defined over hypergeometric products.

\medskip

We note that this class of indefinite nested sums defined over hypergeometric products contains as special cases all iterative sums that arose so far in QCD calculations. In particular, it contains harmonic sums~\cite{Vermaseren:1998uu,Blumlein:1998if} such as	
$$S_{2,1}(n)=\sum_{i=1}^n\frac{1}{i^2}\sum_{j=1}^i\frac{1}{j},$$
generalized harmonic sums~\cite{Moch:02,Ablinger:2013cf} such as
$$S_{1,1}(2,\tfrac12,n)=\sum_{k=1}^n\frac{2^k}{k}\sum_{i=1}^k \frac{\displaystyle(\tfrac12)^{i}}{i}\sum_{j=1}^i \frac{\displaystyle S_1(j)}{j},$$
cyclotomic harmonic sums~\cite{Ablinger:2011te} such as			
$$CS_{(2,1,2),(1,0,2),(2,1,1)}(n)=\sum_{k=1}^n {\frac{1}{(2 k+1)^2}}\sum_{j=1}^k \frac{1}{j^2}\sum_{i=1}^j \frac{1}{2 i+1},$$
binomial sums~\cite{EpExpansionMethods,Ablinger:2014bra} such as
$$\sum_{j=1}^n  
			\frac{4^{j} S_1({j-1})}
			{\binom{2j}{j} j^2}\hspace{1cm}$$
or generalized binomial sums~\cite{Ablinger:2018brx,Ablinger:2020snj} such as
$$\sum_{k=1}^n (\tfrac14)^{k} \big(
			1-{\eta}\big)^k \binom{2 k}{k} 
			\sum_{j=1}^k \frac{4^j}{j^2\binom{2 j}{j}}$$
where the extra parameter $\eta$ encodes the ratio of two masses. More generally, sums of the form
$$\sum_{k=1}^n\left(\prod_{i=1}^k\frac{1+i+i^2}{i+1}\right)\sum_{j=1}^k\frac1{j\binom{4j}{3j}^2}$$
are covered where $f(k)=\prod_{i=1}^k\frac{1+i+i^2}{i+1}$ is hypergeometric in $k$, i.e., its shift-quotient is a rational function in $k$: $$\frac{f(k+1)}{f(k)}=\frac{k^2+3 k+3}{k+2}\in\QQ(k).$$

Similarly, one may activate the package \texttt{HarmonicSums} based on~\cite{IterativeInt,dAlembert,AlgebraicRelations,Ablinger:2017Mellin} to tackle the following problem.

\medskip

\noindent\textbf{GIVEN} a linear differential equation
\begin{equation}\label{Equ:DE}
b_0(x)f(x)+\dots+b_{\lambda}(x)D_x^{\lambda}f(x)=0
\end{equation}
with polynomials $b_0(x),\dots,b_{\lambda}(x)\in\KK[x]$ and initial values, say $f(0),\dots,D_x^{\lambda-1}f(x)|_{x=0}\in\KK$.\\
\textbf{DECIDE} constructively if $f(x)$ can be expressed in terms of iterated integrals defined over hyperexponential functions.

As in the difference equation case, this solver contains as special cases all solutions in terms of iterated integrals that arose so far in QCQ-calculations. Namely, it covers
harmonic polylogarithms~\cite{Remiddi:1999ew} such as	
$$H_{1,-1}(x)=\int _0^x\frac{1}{1-\tau _1}\int _0^{\tau_1}\frac{1}{1+\tau _2}d\tau _2d\tau _1,$$
generalize harmonic polylogarithms~\cite{Moch:02,Ablinger:2013cf} such as
$$H_{2,-2}(x)=\int _0^x\frac{1}{2-\tau_1}\int_0^{\tau_1}\frac{1}{2+\tau_2}d\tau_2d\tau_1,$$
cyclotomic harmonic polylogarithms~\cite{Ablinger:2011te} such as
$$\int _0^x\frac{1}{1+\tau _1+\tau _1^2}\int _0^{\tau_1}\frac{1}{1+\tau_2^2}d\tau _2d\tau_1,$$
radical integrals~\cite{Ablinger:2014bra} such as		
$$\int_0^x\frac{1}{\sqrt{1+\tau_1}}\int_0^{\tau_1}\frac{1}{1+\tau_2}d\tau _2d\tau_1,$$
or generalized radical integrals~\cite{Ablinger:2018brx,Ablinger:2020snj} such as
$$\int _0^x\frac{1}{1-\tau _1+\eta  \tau _1}\int _0^{\tau _1}\sqrt{1-\tau _2} \sqrt{1-\tau _2+\eta  \tau _2}d\tau _2d\tau _1$$
where $\eta$ is the ration of two masses. A more general example is
$$\int_0^x e^{\int_1^{\tau_1}\frac{1}{1+y+y^2}dy}\int_0^{\tau_1}\frac1{1+\tau_2}d\tau_2 d\tau_1$$
where $f(\tau_1)=e^{\int_1^{\tau_1}\frac{1}{1+y+y^2}dy}$ is hyperexponential, i.e. its logarithmic derivative is a rational function in $\tau_1$: $$\frac{D_{\tau_1}f(\tau_1)}{f(\tau_1)}=\frac{1}{1+\tau_1+\tau_1^2}\in\QQ(\tau_1).$$
We remark that \texttt{HarmonicSums} contains also Kovacic's algorithm~\cite{Kovacic:86,Ablinger:2017Mellin} that enables one to find Liouvillian solutions of second order linear differential equations. As a consequence one can find also expressions in terms of iterative integrals over double routed radicals such as $\sqrt[3]{1-\sqrt{1+\tau _1}}$.

\section{Extracting hypergeometric structures from differential equations}\label{Sec:HypStructure}

Given a linear differential equation, one may succeed in finding sufficiently many linearly independent solutions in terms of iterative integrals over hyperexponential functions by using the algorithms given in~\cite{IterativeInt,Kovacic:86,dAlembert,Ablinger:2017Mellin}. But in most cases such a representation is not possible. If the underlying parameter integral, say $f(x,\ep)$, depends on the dimensional parameter $\ep$, one may use refined methods described in~\cite{Ablinger:21} (and inspired by~\cite{BKSS:12}) to find closed form representations of the $\ep$-expansion
$$f(x,\ep)=\phi_l(x)\ep^l+\phi_{l+1}(x)\ep^{l+1}+\dots+\phi_{r}(x)\ep^{r}+O(\ep^{r+1})$$
for some $l,r\in\ZZ$ with $l\leq r$. More precisely, given the differential equation and appropriate initial values, one can decide algorithmically if the coefficients $\phi_l(x),\phi_{l+1}(x),\dots,\phi_r(x)$ can be represented in terms of iterative integrals over hyperexponential functions. But in the most general case (in particular, for more complicated Feynman integrals), also here the available methods will fail.

However, if one searches for representations in terms of hypergeometric structures, the holonomic machinery~\cite{HOLONOMIC} in combination with the available recurrence solver of \texttt{Sigma} might lead to the desired result.

\subsection{Ordinary linear differential equations}\label{Sec:OrdDE}

In the ordinary case, the holonomic approach can be stated as follows. Let 
\begin{equation}\label{Equ:UnivariateSeriesAnsatz}
f(x)=\sum_{n=0}^{\infty} F(n)x^n
\end{equation}
be a (formal) power series. Then there exist $b_0(x),\dots,b_{\lambda}(x)\in\KK[x]$ (not all zero) with
\begin{equation}\label{Equ:HoloDE}
b_0(x)f(x)+\dots+b_{\lambda}(x)D_x^{\lambda}f(x)=0
\end{equation}
if and only if there exist $a_0(x),\dots,a_{\delta}(x)\in\KK[x]$ (not all zero) with
\begin{equation}\label{Equ:HoloRE}
a_0(n)F(n)+\dots+a_{\delta}(n)F(n+\delta)=0.
\end{equation}
We emphasize that this correspondence is fully algorithmic.  If the linear difference equation or the linear differential equation is given, one can compute the other equation explicitly. Here will exploit only one direction: we are given a linear differential equation of an unknown power series~\eqref{Equ:UnivariateSeriesAnsatz} and compute the corresponding linear difference equation for the coefficients $F(n)$; this can be accomplished by plugging the ansatz~\eqref{Equ:UnivariateSeriesAnsatz} into the equation and performing coefficient comparison w.r.t.\ $x^n$. If one succeeds in finding a solution of the recurrence~\eqref{Equ:HoloDE}
with \texttt{Sigma} (see Section~\ref{Sec:REDE}), one obtains automatically a (formal) solution of the linear differential equation~\eqref{Equ:HoloRE}.

\medskip

\noindent \textbf{Example 1:} In order to find a power series solution
$$f(x)=\sum_{n=0}^{\infty} F(n)x^n$$
for
\begin{multline*}
-\left(x^4-64 x^3\right) f^{(4)}(x)-2 \left(5 x^3-144 x^2\right) f^{(3)}(x)\\
-\left(25 x^2-208 x\right) f''(x)-(15 x-8) f'(x)-f(x)=0,
\end{multline*}
we compute the recurrence
	$$8 (n+1) (2 n+1)^3 F(n+1)-(n+1)^4 F(n)=0$$
for the coefficients $F(n)$ of the series $f(x)$.
Using \texttt{Sigma} we obtain the solution
$$F(n)=\frac1{\binom{2n}{n}^3}=\frac{(1)_n(1)_n(1)_n(1)_n}{\big(\tfrac{1}{2}\big)_n\big(\tfrac{1}{2}\big)_n\big(\tfrac{1}{2}\big)_n n!}\frac1{64^{n}}.$$
Thus the full set of power series solutions for the given linear differential equation is given by
$$f(x)=c\cdot\sum_{n=0}^{\infty}\frac{x^n}{\binom{2n}{n}^3}=c\cdot\pFq{4}{3}{1,1,1,1}{\tfrac12,\tfrac12,\tfrac12}{\frac{x}{64}},\quad c\in\KK.$$

\medskip

\noindent \textbf{Example 2:} To find power series solutions
$$f(x)=\sum_{n=0}^{\infty} F(n)x^n$$
for
\begin{multline*}
\left(x^6-32 x^5+256 x^4\right) f^{(6)}(x)+\left(23 x^5-528 x^4+2560 x^3\right) f^{(5)}(x)\\
+\left(171 x^4-2552 x^3+6272 x^2\right)
f^{(4)}(x)+2 \left(245 x^3-2002 x^2+1728 x\right) f^{(3)}(x)\\+2 \left(253 x^2-786 x+72\right) f''(x)+4 (35 x-12) f'(x)+4 f(x)=0
\end{multline*}
we compute the underlying recurrence
$$(n+2) (n+1)^3 F(n)-4 (n+2) (2 n+1)^2 (2 n+3) F(n+1)+16 (2 n+1)^2 (2 n+3)^2 F(n+2)=0.$$
Using \texttt{Sigma} we obtain the solutions
$$F(n)=\frac{1}{\binom{2 n}{n}^2}\Big( c_1+c_2S_1({n})\Big)=\frac{(1)_n(1)_n(1)_n}{\big(\tfrac{1}{2}\big)_n\big(\tfrac{1}{2}\big)_n n!}\frac1{16^{n}}\Big(c_1+c_2S_1({n})\Big),\quad c_1,c_2\in\KK;$$
note that we obtained two linearly independent solutions and thus the solution space of the recurrence is completely determined. Thus the full set of power series solutions is given by
$$f(x)=c_1\cdot\pFq{3}{2}{1,1,1}{\tfrac12,\tfrac12}{\frac{x}{16}}+c_2\sum_{n=0}^{\infty} \frac{S_1(n)}{\binom{2n}{n}^2}x^n,\quad c_1,c_2\in\KK.$$

In the two examples above we obtained the full solution set of the underlying linear recurrence and thus could provide all power series solutions of the given linear differential equation. In general, utilizing the algorithms from~\cite{dAlembert,SymbolicSum,SigmaSolver} we can find all power series solutions whose coefficients can be given in terms of indefinite nested sums over hypergeometric products. This toolbox has been utilized in various concrete QCD calculations such as~\cite{QCDCalc}. In particular, we can find all ${}_pF_q$-solutions of the form ${}_pF_q[\dots,h(x)]$ with $h(x)=c\cdot x$ where $c\in\KK^*$. We note further that one can search also for solutions for more complicated arguments, like $h(x)=\frac{x^2(x^2-9)^2}{(x^2+3)^2}$ as elaborated in~\cite{pFqTransf}. Furthermore, one can hunt also for Puiseux series solutions by variants of this method.

\subsection{Partial linear differential equations}\label{Sec:PartialDE}

For the multivariate case the problem can be stated as follows: Find a power series solution
\begin{equation}\label{Equ:MultiSeriesAnsatz}
f(x_1,\dots,x_r)=\sum_{n_1=0}^{\infty}\dots\sum_{n_r=0}^{\infty}F(n_1,\dots,n_r)x_1^{n_1}\dots x_r^{n_r}
\end{equation}
for a partial linear differential equation
$$\sum_{(s_1,\dots,s_{r})\in T} b_{(s_1,\dots,s_{r})}(x_1,\dots,x_{r})D^{s_1}_{x_1}\dots D^{s_{r}}_{x_r}f(x_1,\dots,x_r)=0$$
where the finite structure set  $T\subset\NN^{r}$ and the polynomial coefficients $b_{(s_1,\dots,s_{r})}(x_1,\dots,x_{r})\in\KK[x_1,\dots,x_r]$ are given.

Already the problem to decide if there is a polynomial solution $f(x_1,\dots,x_r)\in\KK[x_1,\dots,x_r]$ is algorithmically unsolvable. More precisely, it has been shown in~\cite{AP:12} that a solution to this problem is equivalent to solve Hilbert's 10th problem (which is unsolvable). Applying the holonomic translation mechanism~\cite{HOLONOMIC}, i.e., plugging in the ansatz~\eqref{Equ:MultiSeriesAnsatz} into the equation and performing coefficient comparison w.r.t.\ $x_1^{n_1}\dots x_r^{n_r}$, one gets a partial linear difference equation of the form
$$\sum_{(s_1,\dots,s_r)\in S}a_{(s_1,\dots,s_r)}(n_1,\dots,n_r)F(n_1+s_1,\dots,n_r+s_r)=0$$
with a finite structure set $S\subset\ZZ^{r}$ and polynomial coefficients $a_{(s_1,\dots,s_r)}(n_1,\dots,n_r)\in\KK[n_1,\dots,n_r]$. 
However, as for the differential case, it has been shown in~\cite{AP:12} that in general one cannot decide algorithmically if there exists a polynomial solution $F(n_1,\dots,n_r)\in\KK[n_1,\dots,n_r]$. But in the discrete case there are at least methods available~\cite{PLDESolver}
that enable one to hunt for rational solutions. Recently, these techniques have been extended in~\cite{HypStructure} to search for solutions in terms of iterative sums defined over hypergeometric products provided that one can predict the set of sums that arises.

For instance, given the partial linear difference equation
	\begin{multline*}
	\hspace*{-0.3cm}(n+1)^2
	\left(k+n^2+2\right)
	\left(3 k n^2-4 k^2-5 k
	n-12 k+2 n^3+2 n^2-8
	n-8\right) {\color{blue}F(n,k+1)}\\
	+(n+1)^2
	\left(k+n^2+3\right)
	\left(2 k^2-2 k n^2+2 k n+6
	k-n^3-n^2+4 n+4\right){\color{blue}F(n,k+2)}\\
	+(n+1)^2 (k+n+1)
	\left(2 k-n^2+n+4\right)
	\left(k+n^2+1\right)
	{\color{blue}F(n,k)}\\
	-(k+1) n^2 (n+2)^2  
	\left(k+n^2+2 n+2\right){\color{blue}F(n+1,k)}\\
	+k n^2 (n+2)^2
	\left(k+n^2+2 n+3\right)
	{\color{blue}F(n+1,k+1)}=0
	\end{multline*}
and given the predicted set of sums $W=\{S_1(k),S_1(n+k),S_{2,1}(n+k)\}$ together with a degree bound $b=5$ for the maximal total degree of the arising objects in the numerator, one can compute $37$ solutions
$$\frac{p}{(1 + n)^2 (1 + k + n^2)}$$
with 
\begin{align*}
p\in\Big\{&1
		+\frac{1}{2} n S_1({k+n})
		,k,n,k n,k n^2,k n^3,k n^4,k S_1({n}),k n S_1({n}),k n^2 S_1({n}),k n^3 S_1({n}),k S_1({n})^2,\\
		&k n S_1({n})^2,k n^2 S_1({n})^2,k n S_1({n})^3,k S_1({n})^4,k S_{2,1}({n}),k n S_{2,1}({n}),k n^2 S_{2,1}({n}),k n^3 S_{2,1}({n}),\\
		&k S_1({n}) S_{2,1}({n}),k n S_1({n}) S_{2,1}({n}),k n^2 S_1({n}) S_{2,1}({n}),k S_1({n})^2 S_{2,1}({n}),k n S_1({n})^2 S_{2,1}({n}),\\
		&k S_1({n})^3 S_{2,1}({n}),k S_{2,1}({n})^2,k n S_{2,1}({n})^2,k n^2 S_{2,1}({n})^2,k S_1({n}) S_{2,1}({n})^2,k n S_1({n}) S_{2,1}({n})^2,\\
		&k S_1({n})^3,k S_1({n})^2 S_{2,1}({n})^2,k S_{2,1}({n})^3,k n S_{2,1}({n})^3,k S_1({n}) S_{2,1}({n})^3,k S_{2,1}({n})^4\Big\}
		\end{align*}
by using the package \texttt{SolvePLDE}~\cite{HypStructure}.
In particular, one can search for solutions with hypergeometric contributions. In particular, the techniques from~\cite{BKSS:12} have been incorporated in this package to search for the coefficients of $\ep$-expansions.

\subsection{Coupled systems of partial linear differential equations}\label{Sec:SystemPartialDE}
 
Solving coupled system of partial linear differential (and difference) equations is a widely open research topic.
However, as elaborated in~\cite{HypStructure} one succeeds in finding hypergeometric structures for at least some interesting special cases.
For instance, given the system
\begin{align*}
	(x-1) y {\color{blue}D_{x y}f(x,y)}+(x (2 \varepsilon +\tfrac{7}{2})
	-\varepsilon +1){\color{blue}D_x f(x,y)}&\\
	+(x-1) x {\color{blue}D_x^2f(x,y)}+y (2 \varepsilon +1){\color{blue}D_yf(x,y)}+\tfrac{3}{2} (2 \varepsilon +1){\color{blue}f(x,y)}&=0,\\[0.2cm]
	x (y-1) {\color{blue}D_{x y}f(x,y)}+x (4-\varepsilon ) {\color{blue}D_xf(x,y)}+(y-1) y {\color{blue}D_y^2f(x,y)}&\\
	+ (y (\tfrac{13}{2}-\varepsilon
	)-\varepsilon +1){\color{blue}D_yf(x,y)}
	+\tfrac{3 (4-\varepsilon )}{2}{\color{blue}f(x,y)}&=0,
\end{align*}
we can find a power series solution
$$f(x,y)=\sum_{n=0}^{\infty}\sum_{m=0}^{\infty}F(n,m)x^ny^m$$
as follows.
Plugging in this ansatz and comparing coefficients w.r.t.\ $x^ny^n$ yields the first-order coupled system
\begin{align*}
		\tfrac{3}{2} (2 \varepsilon +1) F(n,m)-n (\varepsilon -1) F(n+1,m)&=0,\\
		-\tfrac{3}{2} (\varepsilon -4) F(n,m)-m (\varepsilon -1)F(n,m+1)&=0.
\end{align*}
In general, the Ore--Sato theorem~\cite{AP:02} provides a criterion for such first-order systems whether a solution can be given in terms of Gamma-functions and Pochhammer symbols. In~\cite{HypStructure} a surprisingly simple method has been elaborated and implemented in the package \texttt{HypSeries} that enables one to produce such solutions in terms of hypergeometric products. In our concrete example we compute the solution
\begin{align*}
F(n,m)=&\Big(
		\prod_{i=1}^n \frac{(1+2 i) (3
			+i
			-\varepsilon 
			)}{2 i (-2
			+i
			+\varepsilon 
			)}\Big) \prod_{i=1}^m \frac{(1
			+2 i
			+2 n
			) (i
			+2 \varepsilon 
			)}{2 i (-2
			+i
			+n
			+\varepsilon 
			)}=
		\frac{\big(
		\frac{3}{2}\big)_{m
			+n
		} (4-\varepsilon )_n (1+2 \varepsilon )_m}{m! n! (-1+\varepsilon )_{m
			+n}}
\end{align*}
and thus the hypergeometric series solution
	\begin{align*}
	f(x,y)=\sum_{n=0}^{\infty}\sum_{m=0}^{\infty}\frac{\big(
				\frac{3}{2}\big)_{m
					+n
				} (4-\varepsilon )_n (1+2 \varepsilon )_m}{m! n! (-1+\varepsilon )_{m
					+n
			}}
	\end{align*}
which is closely related to Appell-like structures.
Expanding the summand in $\ep$ gives
	$$f(x,y)=\ep^{-1}\sum_{n=0}^{\infty}\sum_{m=0}^{\infty}F_{-1}(n,m)+\ep^{0}\sum_{n=0}^{\infty}\sum_{m=0}^{\infty}F_{0}(n,m)+\dots$$
with
\begin{align*}
		F_{-1}(n,m)&=-\frac{1}{6} \frac{x^m y^n (3+n)! \big(
			\frac{3}{2}\big)_{m
				+n
		}}{n! (-2
			+m
			+n
			)!}\\
		F_{0}(n,m)&=
		\bigg[\dots
		6 S_1({n})
		+6 S_1({m+n})-12 S_1({m})
		\bigg] 
		\frac{x^m y^n (3+n)! \big(
			\frac{3}{2}\big)_{m
				+n
		}}{n! (-2
			+m
			+n
			)!}.
		\end{align*}
Finally, using the summation package~\texttt{Sigma} with its summation algorithms~\cite{SymbolicSum} one obtains	
$$\sum_{n=0}^{\infty}\sum_{m=0}^{\infty}F_{-1}(n,m)=-\frac{15 x^6 }{4 (x-y)^4(1-x)^{7/2}}
			-\frac{15 y^3Q(x,y)}{64 (x-y)^4 (1-y)^{13/2}}$$
for some polynomial $Q(x,y)$. For further details, in particular for the result of the constant coefficient $\sum_{n=0}^{\infty}\sum_{m=0}^{\infty}F_{0}(n,m)$ we refer to~\cite{HypStructure}.

\section{Conclusion}\label{Sec:Conclusion}
We illustrated up-to-date computer algebra methods and implementations that assist the user to find hypergeometric structures for various classes of linear differential equations. While the ordinary case (see Sections~\ref{Sec:REDE} and~\ref{Sec:OrdDE}) has been pushed forward non-trivially in various directions, the partial case (see Sections~\ref{Sec:PartialDE} and~\ref{Sec:SystemPartialDE}) has been neglected so far. We hope that the recent results from~\cite{HypStructure} and the supplementary material of this article will open up new developments for improved differential equation solvers that are instrumental for future explorations and calculuations of parameter Feynman integrals.

\vspace{5mm}\noindent
{\bf Acknowledgment.}~
This work was supported by the European Union's Horizon 2020 research and innovation programme under the Marie 
Sk\l{}odowska--Curie grant agreement No. 764850, SAGEX and the Austrian Science Fund (FWF) 
grants SFB F50 (F5009-N15) and P33530.

\scriptsize

\setlength{\bibsep}{0pt plus 0.3ex}

\end{document}